# Title: Electrically Programmable Correlated Topology and Magnetism in a Moiré Trilayer

**Authors:** Christiano Wang Beach[1]*, Courtney Baier[2]*, Kaijie Yang[3], Huiyuan Zheng[3], Yueyao Fan[3], Weijie Li[1], Shuai Yuan[1], Yifan Zhao[1], Yue Sun[1], Chaowei Hu[1], Takashi Taniguchi[4], Kenji Watanabe[5], Jiun-haw Chu[1], Liang Fu[6], Ting Cao[3], Satoshi Okamoto[7], Di Xiao[3,1#], Xiaodong Xu[1,3#]

[1]Department of Physics, University of Washington, Seattle, Washington 98195, USA
[2]Department of Chemistry, University of Washington, Seattle, Washington 98195, USA
[3]Department of Materials Science and Engineering, University of Washington, Seattle, Washington 98195, USA
[4]Research Center for Materials Nanoarchitectonics, National Institute for Materials Science, 1-1 Namiki, Tsukuba 305-0044, Japan
[5]Research Center for Electronic and Optical Materials, National Institute for Materials Science, 1-1 Namiki, Tsukuba 305-0044, Japan
[6]Department of Physics, Massachusetts Institute of Technology, Cambridge, Massachusetts 02139, USA
[7]Materials Science and Technology Division, Oak Ridge National Laboratory, Oak Ridge, Tennessee 37831, USA

[#]Corresponding author's email: dixiao@uw.edu; xuxd@uw.edu

**Abstract: Strong electron–electron interactions underlie a wide range of quantum many-body phenomena, including magnetism, superconductivity, and charge fractionalization. A central goal is to achieve *in situ* control over lattice geometry, bandwidth, and band topology within a single platform. Here we realize such an electrically programmable quantum many-body system in an alternating twisted trilayer $MoTe_2$, where an out-of-plane displacement field continuously modifies the layer polarization, effective lattice, and topology of the moiré bands. At zero displacement field, the system realizes a triangular lattice hosting a correlated insulator at one hole per moiré unit cell ($\nu = -1$). Doping this state produces strongly asymmetric magnetic responses: double-exchange-like ferromagnetism for $|\nu| > 1$, and signatures of spin polarons and antiferromagnetism for $|\nu| < 1$. At large displacement field, interlayer hybridization reconstructs the electronic structure into a honeycomb lattice with a flat Chern band, supporting integer and fractional Chern insulators. Magneto-optical measurements further reveal the signatures of gap closure and Landau-level formation from a spin-polarized Fermi surface near the crossover between the two regimes. These results establish a unified, electrically tunable platform in which correlated magnetism and topological states emerge from a single controllable band structure.**

Moiré superlattices in transition metal dichalcogenides (TMDs) have emerged as a powerful platform for realizing correlated and topological quantum states[1–3]. In particular, twisted bilayer $MoTe_2$ has been shown to host the fractional quantum anomalous Hall effect[4–6] and unconventional superconductivity[7], intertwining strong electron correlations, topology, and magnetism. Extending from bilayer to trilayer structures, as pioneered in graphene moiré systems[8–13], introduces additional layer degrees of freedom that can modify lattice symmetry, electronic bandwidth, and band topology. Recent theoretical works have suggested that twisted TMD multilayers can host flat bands with electrically tunable topology[14–18], including fractional Chern insulator (FCI) states at zero magnetic field. Such systems offer a promising route toward engineering controllable many-body Hamiltonians within a single platform.

Here, we realize an alternating-twisted trilayer $MoTe_2$ ($ATMoTe_2$) with a twist angle of 3.9° and demonstrate that an out-of-plane displacement field continuously tunes the underlying band structure between two distinct regimes. At low displacement field, the system forms a triangular lattice hosting a correlated insulating state at $\nu = -1$, from which doping gives rise to double-exchange ferromagnetism for $|\nu| > 1$, spin polaron behavior and collinear antiferromagnetism for $|\nu| < 1$. At large displacement field, the same system reconstructs into a honeycomb lattice with flat Chern bands, enabling the emergence of robust integer and fractional Chern insulators. These two regimes are connected through a displacement-field-driven band reconstruction, where calculations predict a gap closure associated with a massless Dirac band at a critical displacement field. Experimentally, we optically observe Landau levels of a spin-polarized Fermi surface in the doped correlated insulator below the critical *D* field, providing evidence for band closure and reconstruction under electric-field control.

**Emergent ferromagnetism**

Figure 1a shows the schematic of the moiré superlattice of $ATMoTe_2$, where blue and red dots represent the transition metal atoms in the outer and inner layers, respectively. We define the twist angle between the first and middle layers as $\theta_{12}$, and between the middle and third layers as $\theta_{23}$. In the alternating twisted trilayer structure, the first and third layers are aligned, while the middle layer is twisted at a relative angle $\theta$ (i.e., $\theta_{12} = -\theta_{23} = \theta$). Within each moiré unit cell, there are two inequivalent high-symmetry sites (moiré orbitals), denoted by MXM and XMX, where M and X represent Mo and Te atoms, respectively. Charge transfer between layers generates a permanent electric dipole. For the XMX configuration, the dipoles from the top and bottom layers point toward the middle layer, whereas for MXM they point outward. As a result, the moiré potential in the middle layer forms a triangular lattice with enhanced strength, favoring strongly interacting flat-band physics[14–16,18].

Tuning the electric field can transition the system between a triangular lattice at low electric fields and a honeycomb lattice at large electric fields. The left panel of Figure 1b shows the layer-resolved density of states for the undoped system at zero displacement field ($D/\epsilon_0$ = 0 V/nm). The highest-energy band has a narrow bandwidth and is predominantly localized in the middle layer, whereas the second band is much more dispersive and has mainly top- and bottom-layer character. Consequently, upon hole doping into the highest band, the carriers effectively occupy a triangular lattice formed by orbitals in the middle layer. As displacement field tunes the interlayer potential, at sufficiently large *D*, the highest bands acquire strongly hybridized middle- and bottom-layer character, while the top-layer states are shifted to lower energies (Fig. 1b right panel). The doped holes therefore experience an effective honeycomb lattice, whose two moiré sublattices reside in the middle and bottom layers, respectively. Reversing the electric fields will swap the roles between the top and bottom layers.

We probe magnetic properties using reflective magnetic circular dichroism[19] (RMCD) at 1.6 K (see Methods). Figure 1c shows RMCD signal as a function of filling factor $\nu$ (carriers per moiré unit cell) and displacement field $D$. A highly tunable ferromagnetic phase diagram is observed, with two strongly ferromagnetic regions centered around $D/\epsilon_0 \sim \pm 0.34$V/nm, connected by a continuous stripe as $D$ is swept (see similar data from a second device in Extended Data Fig. 1). Figure 1d shows a zoomed-in plot of RMCD near $D/\epsilon_0 = -0.34$ V/nm. The RMCD phase diagram exhibits the characteristic "Y"-shaped pattern with enhanced critical displacement fields near $\nu = -1$ and $\nu = -2/3$.[20]

To confirm ferromagnetism, we perform magnetic hysteresis measurements at selected fillings for $D/\epsilon_0 = -0.34$ V/nm (see insets in Fig. 1c). The hysteretic component, ΔRMCD, is plotted in Fig. 1e as a function of $\nu$, showing enhanced coercive fields $\mu_0 H_C$ near $\nu = -1$ (~100 mT) and $\nu = -2/3$ (~20 mT). The ferromagnetism is robust, with Curie temperatures of approximately 11 K and 3.5 K at $\nu = -1$ and $\nu = -2/3$, respectively (Fig. 1f). These values are comparable to those reported in twisted bilayers[20]. We can see that the robust magnetic behavior at large displacement field resembles that of twisted bilayer, consistent with the reconstruction into a honeycomb lattice at large displacement field. Extended Data Fig. 2a highlights Δ RMCD as a function of displacement field at $\nu \approx -1$. The hysteresis loop narrows abruptly as $D$ approaches zero, while evolving more gradually at larger $|D|$. This asymmetry highlights the transition between distinct magnetic regimes, reflecting the underlying electric-field-driven change in lattice geometry from triangular to honeycomb.

**Gate tunable integer and fractional Chern insulators**

The strong similarity of the magnetic response between the large-$D$ regime in AT$MoTe_2$ and twisted bilayers suggests that the reconstructed honeycomb-lattice regime may also host nontrivial topology. To investigate this possibility, we perform trion sensing measurements, following approaches established in twisted bilayer systems[20]. We observe pronounced exciton and trion photoluminescence in AT$MoTe_2$ (Extended Data Fig. 3a), confirming a direct-gap electronic structure. Figure 2a shows the spectrally integrated trion intensity as a function of filling and displacement field. Suppression of trion intensity at specific fillings indicates the formation of incompressible states. The resulting map reveals multiple incompressible features with distinct topological character, as discussed below.

We first examine the incompressible states at large $D$. Figure 2b shows the PL intensity plot as a function of $\nu$ and photon energy at $D/\epsilon_0 = -0.34$ V/nm. Trion PL intensity dips and peak shifts are evident at $\nu = -1$ and $-2/3$, and a faint feature at $-3/5$. Comparison of Fig. 1c and Fig. 2a reveals that these incompressible states are in the phase space of ferromagnetism, i.e. they have spontaneous time reversal symmetry breaking. We further perform PL measurements as a function of magnetic field. Figure 2c is the optically detected fan diagram, resulting from spectrally integrated trion PL as a function of filling and magnetic fields. The density of incompressible states at $\nu = -1$, $-2/3$, and $-3/5$ shifts as magnetic field increases. Comparison with the Streda formula identifies three Chern insulator states $(-1, -1)$, $(-2/3, -2/3)$ and $(-3/5, -3/5)$, with $(C, \nu)$ as Chern number and filling factor. The observation of integer and fractional Chern insulator state at zero magnetic field further supports our understanding that the system at large $D$ field is a honeycomb lattice with flat Chern bands[21,22].

The system becomes topologically trivial at $D/\epsilon_0$ = 0. The measurement of PL as a function of doping, shown in Fig. 2d, reveals an incompressible state at $\nu = -1$, but not at −2/3 and −3/5. The incompressible state can be understood as a correlated insulator with a full filling of a triangular lattice[1,3]. Indeed, optically detected fan diagrams show that the $\nu = -1$ state is not dispersive as a function of magnetic field, confirming its topologically trivial character (Fig. 2e). The fan diagram measurement also suggests an additional field-induced incompressible state at $\nu = -2/3$ and $\nu = -1/2$. These states are also topologically trivial, likely a charge ordered state.

The transition between the topologically trivial triangular lattice at $D/\epsilon_0 = 0$ and the Chern band regime at large $D$ is driven by an electric-field-controlled band reconstruction[18]. Using the $\nu = -1$ state as an example, Fig. 2a shows that the trion PL dip persists as $D$ increases, then disappears near a critical displacement field (indicating gap closure), and subsequently re-emerges at larger $D$ in the $C = -1$ Chern insulator state. Taking a linecut of this phase diagram at $\nu = -1$ (Extended Data Fig. 3b) further supports the closing and reopening of the gap. This evolution demonstrates electric-field control of the band topology through tunable layer polarization and hybridization.

This topological phase transition is also supported by band structure calculations. Extended Data Fig. 4 shows a transition from a trivial band ($C = 0$) at $D/\epsilon_0 = 0$ to a flat Chern band ($C = -1$) at large $D$, with a Dirac band emerging at the critical field. Experimentally, we observe Landau level formation. The bottom panel of Fig. 2a shows the spectrally integrated PL at magnetic field of 8T. For $|\nu| > 1$, *i.e.* doped correlated insulator states, PL intensity shows stripe pattern versus doping for $D$ field smaller than the critical value, a manifestation of the formation of Landau levels.[23–25] Extended Data Fig. 5 also shows the spectrally integrated PL intensity as a function of doping and magnetic field at fixed $D/\epsilon_0$ = −0.13 V/nm. Landau fan features originating from the doped correlated insulator at $\nu = -1$ are observed[26]. These results support the presence of a spin-polarized Fermi surface, which provides evidence for band closure and reconstruction near the critical displacement field.

**Double-exchange ferromagnetism for $|\nu| > 1$**

The measurements above establish that large-*D* produces a honeycomb-lattice with flat Chern-bands. We now return to the low-*D* triangular-lattice regime and examine how doping the correlated insulator at $\nu = -1$ generates sharply asymmetric magnetic states. Comparison of Fig. 1c and Fig. 2a at 8T shows that the ferromagnetic phase coincides with the presence of a Fermi surface. This indicates the formation of a spontaneous spin-polarized Fermi surface upon doping the correlated insulating state in the triangular lattice. To further establish the ferromagnetism and the underlying mechanism, we focus on the magnetic response at $D/\epsilon_0 = 0$. Figure 3a plots RMCD at cycling magnetic fields for selected filling factors near $\nu = -1$. The emergence of magnetic hysteresis immediately upon doping beyond $\nu = -1$ unambiguously demonstrates spontaneous ferromagnetism in the over doped regime.

We extract the hysteric component of RMCD signal, ΔRMCD, and plot it as a function of doping. As shown in Fig. 3b, ΔRMCD exhibits a dome-like shape and is highly asymmetric with respect to electron and hole doping of the $\nu$ = -1 correlated insulator. The RMCD hysteresis only emerges at doping slightly above $\nu$ = -1 and vanishes for the underdoped regime. The coercive field initially increases with doping, peaks near $\nu$ = -1.2, and then decreases upon further doping. We also performed temperature-dependent RMCD at zero magnetic field. The Curie temperature exhibits a similar dome-like dependence, increasing with doping, reaching a maximum of ~3.5 K near $\nu \approx -1.2$ (Fig. 3c), and decreasing at higher doping.

In a triangular lattice with strong electron–electron interactions, at $\nu = -1$ the system likely forms a frustrated 120 degree frustrated antiferromagnetic state (Fig. 3d)[27,28]. We performed RMCD measurements at $\nu = -1$ as a function of temperature (Extended Data Fig. 2c). We then use the slope of RMCD at $\mu_0 H$ = 0 as an approximation of magnetic susceptibility $\chi_{MCD}$.[1,29] The Curie-Weiss fit of 1/ $\chi_{MCD}$ as function of temperature (Fig. 3d) yields a nearly vanishing Curie-Weiss temperature. This indicates a very weak exchange interaction between the local moments.

Upon doping the insulating state, the extra hole carriers can mediate interactions between the local moments in the middle layer, resulting in the observed ferromagnetism. Figure 3d plots Curie temperature, and normalized remanent RMCD, coercive fields, and saturated RMCD, as a function of $\nu$. Remarkably, these four independent observables exhibit the same doping dependence. They increase with $\delta$ ($\delta = |\nu| - 1$), peak at the same doping, and then decrease with higher doping. This observation demonstrates that these observables are governed by a common underlying energy scale that sets the intrinsic stability of the ferromagnetic state.

Two kinetic mechanisms may, in principle, account for this behavior. If the doped holes enter the same strongly correlated middle-layer band that hosts the local moments, their motion can favor a Nagaoka-like ferromagnetic state[30,31] (more broadly a Stoner ferromagnetic state[32]). Alternatively, if the holes predominantly occupy itinerant states associated with the outer layers while remaining exchange-coupled to the middle-layer moments, ferromagnetic alignment can enhance their mobility through a double-exchange-like mechanism[33,34]. The dome-shaped doping dependence in Fig. 3d alone does not distinguish between these two scenarios.

To identify the relevant regime, we use energy- and layer-sensitive exciton spectroscopy to determine the layer character of the doped carriers. At $D/\epsilon_0$ = 0, we find that the added holes reside predominantly in the top- and bottom-layer states rather than in the middle-layer Hubbard band (Extended Data Fig. 6). This observation rules out Nagaoka ferromagnetism since it is single orbital phenomena. Instead, a double-exchange-like mechanism is favored, in which itinerant outer-layer holes mediate ferromagnetic coupling between middle-layer local moments. This also implies that the lowest-energy charged excitations involve charge transfer between distinct layer sectors. This suggests the correlated insulator at $\nu = -1$ is a layer-selective charge-transfer insulator.

**Spin Polaron and collinear antiferromagnet for $|\nu|$ < 1**

Reducing the hole density below the $\nu = -1$ correlated insulator produces a qualitatively different magnetic response from that observed for $|\nu| > 1$. We observe distinct magnetic responses in different doping regimes under applied magnetic fields. To track these magnetic phases across doping regimes, we define a derivative magnetic circular dichroism signal d(MCD)/dE as the difference between $\sigma^+$ and $\sigma^-$ detection of d($\Delta R/R_0$)/dE under linearly polarized excitation (see Methods). Figure 4a shows d(MCD)/dE at $\mu_0 H = 0$ T (top) and 0.5 T (bottom), with their difference presented in Extended Data Fig. 7a. The d(MCD)/dE signal in the ferromagnetic regime shows little change with applied magnetic field. In contrast, pronounced field-dependent d(MCD)/dE responses are observed for $\nu$ slightly below -1 and near $\nu \approx -1/2$, while the response remains weak near $\nu = -2/3$. These distinct behaviors indicate the presence of different magnetic states for $|\nu| < 1$. In the following, we focus on $D/\epsilon_0 = 0$, where the system realizes an ideal triangular lattice.

We first examine the magnetic response for $\nu$ slightly below -1. d(MCD)/dE measurements as a function of magnetic field reveal a non-monotonic dependence on $\mu_0 H$ (Fig. 4b). The signal initially increases, exhibits a plateau at intermediate fields, and then increases again toward saturation at higher fields. This behavior is reminiscent of observations in $MoTe_2/WSe_2$ heterobilayers and has been attributed to spin-polaron formation[35,36]. In this picture, doping an electron into a triangular hole insulator leads to the formation of a bound state between the electron and a hole spin excitation, resulting in a spin-polaron with total spin $3/2$, which gains kinetic energy *t* through correlated hopping[35].

The spin-polaron scenario naturally explains the observed d(MCD)/dE behavior. At low magnetic field, spins gradually align, leading to an increase in d(MCD)/dE. Once the field exceeds a critical saturation value $\mu_0 H_s$, the background spins become fully polarized, and the presence of spin polarons reduces the total magnetization, giving rise to a plateau. Then, at a higher magnetic transition field $\mu_0 H_m$, Zeeman energy overcomes the spin-polaron binding, restoring full spin polarization and leading to a second saturation.

To further test this picture, we examine the ratio between the plateau and full saturation d(MCD)/dE signals. The expected ratio is $(1 - 3|\delta|)/(1 - |\delta|)$, which agrees well with the experimental data (Fig. 4c). In addition, the plateau feature weakens at elevated temperature (Extended Data Fig. 6b), consistent with a finite binding energy[36]. Taken together, these observations are consistent with spin-polaron formation in the doped triangular lattice.

We next examine the state near $\nu = -2/3$, which shows weak d(MCD)/dE response at low magnetic field. From temperature-dependent RMCD measurements, we extract the magnetic susceptibility $\chi_{\mathrm{MCD}}$ from the slope near $\mu_0 H = 0$. Figure 4d shows $1/\chi_{\mathrm{MCD}}$ as a function of temperature. A Curie–Weiss fit yields a Curie–Weiss temperature of approximately $-2.5$ K, indicating antiferromagnetic interactions at the base temperature of 1.6 K. The absence of hysteresis and zero-field RMCD further supports this interpretation. At $\nu = -2/3$, corresponding to one hole removed from the triangular lattice, the charge configuration forms a honeycomb structure. A collinear antiferromagnetic ground state is naturally expected, where spins at adjacent site are anti-aligned with each other[37].

In summary, we establish $\mathrm{ATMoTe_2}$ as a highly tunable quantum many-body platform where an out-of-plane displacement field continuously reprograms the underlying band structure and emergent phases. Figure 4e summarizes the phase diagram. At low displacement field, the system forms a strongly interacting triangular lattice with correlated magnetism arising from the kinetic ferromagnetism. Application of a large displacement field reconstructs it into a honeycomb lattice with flat Chern bands supporting robust integer and fractional Chern insulators. These two regimes are connected through an electric-field-driven band reconstruction, with optical signatures of a spin-polarized Fermi surface and gap closure near the critical point. Our results demonstrate a unified platform where correlated magnetism and interaction-driven topology can be selectively accessed and linked within a single controllable system.

## Methods

### Device fabrication

We first prepared the back gate by sequentially picking up contact graphite, hexagonal boron nitride (hBN) dielectric and bottom graphite flakes using a polycarbonate (PC) dry-transfer technique. We then released them onto the 285 nm $SiO_2$/Si substrates with pre-patterned gold electrodes. The assembled back gate was immersed in chloroform for 10 minutes to dissolve the polymer, followed by AFM cleaning to remove residual PC. To obtain twisted trilayer (ML/ML/ML) $MoTe_2$ heterostructure, we exfoliated high-quality home-grown $MoTe_2$ crystals inside an argon-filled glovebox and identified monolayer $MoTe_2$ flakes. A thin hBN flake was used to pick up the first ML $MoTe_2$ flake, which was rotated to the target twist angle of $4^o$ before picking up the second third of the ML $MoTe_2$ flake. Then the picked-up flakes are rotated to a target twist angle of $-4^o$. The assembled heterostructure was released on the prepared back gate. To ensure clean interfaces, AFM cleaning was employed to expel trapped gas bubbles. This yielded a large homogeneous area. Finally, a graphite top gate and top hBN dielectric were transferred over the stack to complete the device fabrication.

### Optical measurements

All measurements were carried out in closed-loop magneto-optical cryostats (attoDRY 2100XL) with attocube xyz nanopositioners and xy scanners, superconducting magnets (9T z-axis or a vector set of 5T z-axis & 2T x- and y-axis), and a base temperature of 1.6 K.

Reflectance contrast and PL measurements were taken with linearly polarized excitation from a fiber-coupled halogen lamp and 632.8 nm HeNe laser, respectively. The beam was focused on the samples using a high-NA nonmagnetic cryogenic objective, yielding a spot size of about 1 μm. Reflectance (PL) was collected by the same objective and analyzed using a quarter-wave plate and linear polarizer to select out right and left circularly polarized components. The signal was then sent to a spectrometer, where the signal is dispersed with a diffraction grating (Princeton Instruments, 600 grooves/mm at 1 μm blaze or 300 grooves/mm at 1.2 um) and detected by a liquid nitrogen cooled infrared InGaAs CCD (Princeton Instruments PyLoN-IR 1.7). The reflectance contrast is defined as $\Delta R/R_0 = (R-R_0)/(R_0-C_{dark})$, where $R$ is the reflected signal from the sample, $R_0$ is the reference signal from the region without $MoTe_2$, and the $C_{dark}$ denotes the dark counts of the InGaAs CCD camera. d($\Delta R/R_0$)/dE is defined as the derivative of reflectance contrast with respect to photon energy $E$. The derivative is taken to enhance the fine spectral features. d(MCD)/dE is defined as the difference between $\sigma^+$ and $\sigma^-$ detection of d($\Delta R/R_0$)/dE at the low energy trilayer resonance.

RMCD measurements were taken with the laser excitation slightly below the trilayer trion resonance. The laser source was a broadband supercontinuum (NKT SuperK Fianium FIR-20), which passed through a homebuilt filter to achieve a narrow excitation bandwidth (~ 1 nm). The out-of-plane magnetization of the sample induces a RMCD signal ΔR, the difference between the right and left-circularly polarized light reflection. To obtain the normalized RMCD, ΔR/R, the laser intensity was chopped at p = 850Hz and the phase was modulated by λ/4 via a photoelastic modulator at f = 50kHz. An InGaAs avalanche photodiode detector was used to collect the signal, and the output was read by a lock-in amplifier (Zurich HF2LI). The ratio between the p-component signal $I_p$ and f-component signal $I_f$ gives the RMCD signal: $\Delta R/R = I_f/(J_1(\pi/2) \times I_p)$ where $J_1$ is the first-order Bessel function.

## Continuum model and DOS calculation

We describe the alternating twisted trilayer $MoTe_2$ using a continuum model[15,38,39]. The basis for the model is defined by the valley $\eta = K, K'$ and layer $l = 1,2,3$ for the top, middle, and bottom layer, respectively. The Hamiltonian in the layer basis by

$$h_\eta^0(r) = \begin{pmatrix} \frac{p^2}{2m} + V_{\eta,1}(r) - \Delta_D & t_{\eta,12}(r) & t_{\eta,13,\gamma} \\ t_{\eta,12}(r)^* & \frac{p^2}{2m} + V_{\eta,2}(r) + V_{\eta,2,\gamma} & t_{\eta,23}(r) \\ t_{\eta,13,\gamma}^* & t_{\eta,23}(r)^* & \frac{p^2}{2m} + V_{\eta,3}(r) + \Delta_D \end{pmatrix}.$$

$m = 0.62 m_e$ is the effective mass of the monolayer $MoTe_2$ with $m_e$ as the bare electron mass. The twist angle is taken as 3.89°. $p$ is the momentum operator. $V_{\eta,l}(r) = \sum_G V_{\eta,l,G} e^{-i\eta G\cdot r}$ , $V_{\eta,2,\gamma}$ are the intralayer moiré potentials, where the reciprocal lattice vectors $G$ are summed over the first harmonic reciprocal lattice vectors $\{\pm b_1, \pm b_2, \pm(-b_1 + b_2)\}$ . $b_1, b_2$ are primitive moiré reciprocal lattice vectors. $t_{\eta,12}(r), t_{\eta,23}(r), t_{\eta,13,\gamma}$ are the interlayer moiré potentials $t_{\eta,12}(r) = \sum_G t_{\eta,12,Q+G} e^{-i\eta(Q+G)\cdot r}$ , $t_{\eta,23}(r) = \sum_G t_{\eta,23,Q+G} e^{-i\eta(-Q-G)\cdot r}$ with $G$ summed over $\{\gamma, -b_2, b_1 - b_2\}$. $Q = -b_1/3 + 2b_2/3$. $\Delta_D$ is the potential difference induced by the external displacement field. The independent parameters fitted to the ab initio calculation for the relaxed trilayer systems are $V_{K,2,\gamma} = -4.38$ meV, $V_{K,1,b_1} = V_{K,3,b_1} = 16.3 e^{i\,0.877\pi}$ meV, $V_{K,2,b_1} = 28.78\, e^{-i0.82\pi}$ meV, $t_{K,12,Q} = t_{K,23,Q}^* = -20.64$ meV, and $t_{K,13,\gamma} = -5.16$ meV.

The eigenstates of the Hamiltonian are solved in the plane-wave basis and denoted as $u_{n,\eta,l,G}(k)$ for the band $n$ with the energy $\epsilon_{n,k}$ . The real space wavefunction is $\psi_{n,k,\eta,l}(r) = \sum_G e^{i(k-G-\eta Q_l)\cdot r} u_{n,\eta,l,G}(k)$ with $Q_l = -Q, Q, -Q$ for $l = 1,2,3$ respectively. $Q = b_1/3 + b_2 2/3$ is located on the moire Brillouin zone corner.

The self-consistent calculations based on relaxed moiré structure are performed with SIESTA package[40–42] with scalar-relativistic Optimized Norm-Conserving Vanderbilt Pseudopotential (ONCVPSP)[43], PBE exchange correlation functional[44], double-zeta plus polarization (DZP) basis and a cutoff of 100.0 Ry. Electronic structure is calculated on moiré structure relaxed by MLFF. Spin-orbit coupling (SOC) is added with on-site approach for band structure calculations[45].

## Machine Learning Force Field

We employ a DeePMD[46] machine-learning force field within the LAMMPS package[47] to relax the alternatively twisted trilayer $MoTe_2$ structures. The training data were generated from ab initio molecular dynamics simulations of the 6.01° 1+2 $tMoTe_2$ structure performed using the Vienna Ab initio Simulation Package (VASP)[48], with the projector-augmented wave (PAW) pseudopotentials[49,50] and the Perdew-Burke-Ernzerhof (PBE) exchange-correlation functional[44]. The van der Waals interactions were treated using the DFT-D2 method of Grimme[51]. Full model details are provided in (Fan, 2025)[52]. The atomic configuration of alternatively twisted trilayer $MoTe_2$ structures is relaxed until the maximum atomic force norm is smaller than $5.0\times10^{-3}$ eV/Å.

**Acknowledgements:** This project is mainly supported by the U.S. Department of Energy (DOE), Office of Science, Basic Energy Sciences (BES), under the award DE-SC0018171. The magneto-optical measurements are partially supported by Vannevar Bush Faculty Fellowship (Award number N000142512047). Device fabrication is partially supported by AFOSR FA9550-24-1-0004. Bulk $MoTe_2$ crystal growth and characterization is partially supported by UW MEM-C award number DMR-2308979. T.C. acknowledges the support by the U.S. Department of Energy, Office of Basic Energy Sciences, under Contract No. DE-SC0025327 for part of the theoretical analysis. Y.Z. acknowledges the support from UW Clean Energy Institute Distinguished Postdoctoral Fellowship. D.X. acknowledges the support from DoE BES under the award DE-SC0012509. S.O. acknowledges the support by the U.S. Department of Energy, Office of Science, Basic Energy Sciences, Materials Sciences and Engineering Division. K.W. and T.T. acknowledge support from the CREST (JPMJCR24A5), JST and World Premier International Research Center Initiative (WPI), MEXT, Japan. X.X. acknowledges support from the State of Washington funded Clean Energy Institute and from the Boeing Distinguished Professorship in Physics.

**Author contributions:** X.X. conceived of and supervised the experiment. C.W.B., C.B., and W.L. performed the magneto-optical measurements with assistance from S.Y. and Y.S.  C.B. and C.W.B. fabricated the devices. C.W.B., C.B, W.L., TC, DX, and X.X. analyzed and interpreted the results. Y.F. and T.C. performed first principle calculations. K.Y., H.Z., S.O., T.C., and D.X. fitted the continuum model and analyzed the mechanism of magnetization. L.F. contributed to the spin-polaron interpretation. S.O. contributed to the theoretical discussion on various mechanisms of ferromagnetic states. T.T. and K.W. synthesized the hBN crystals. C.H. and Y.Z. grew and characterized the bulk $MoTe_2$ crystals, supervised by J.H.C. X.X., C.W.B., D.X. and C.B. wrote the paper with input from all authors. All authors discussed the results.

**Competing interests:** The authors declare no competing interests.

**Data availability:** Source data that reproduces the plots are provided with this paper.

### Method References

## Main Figures:

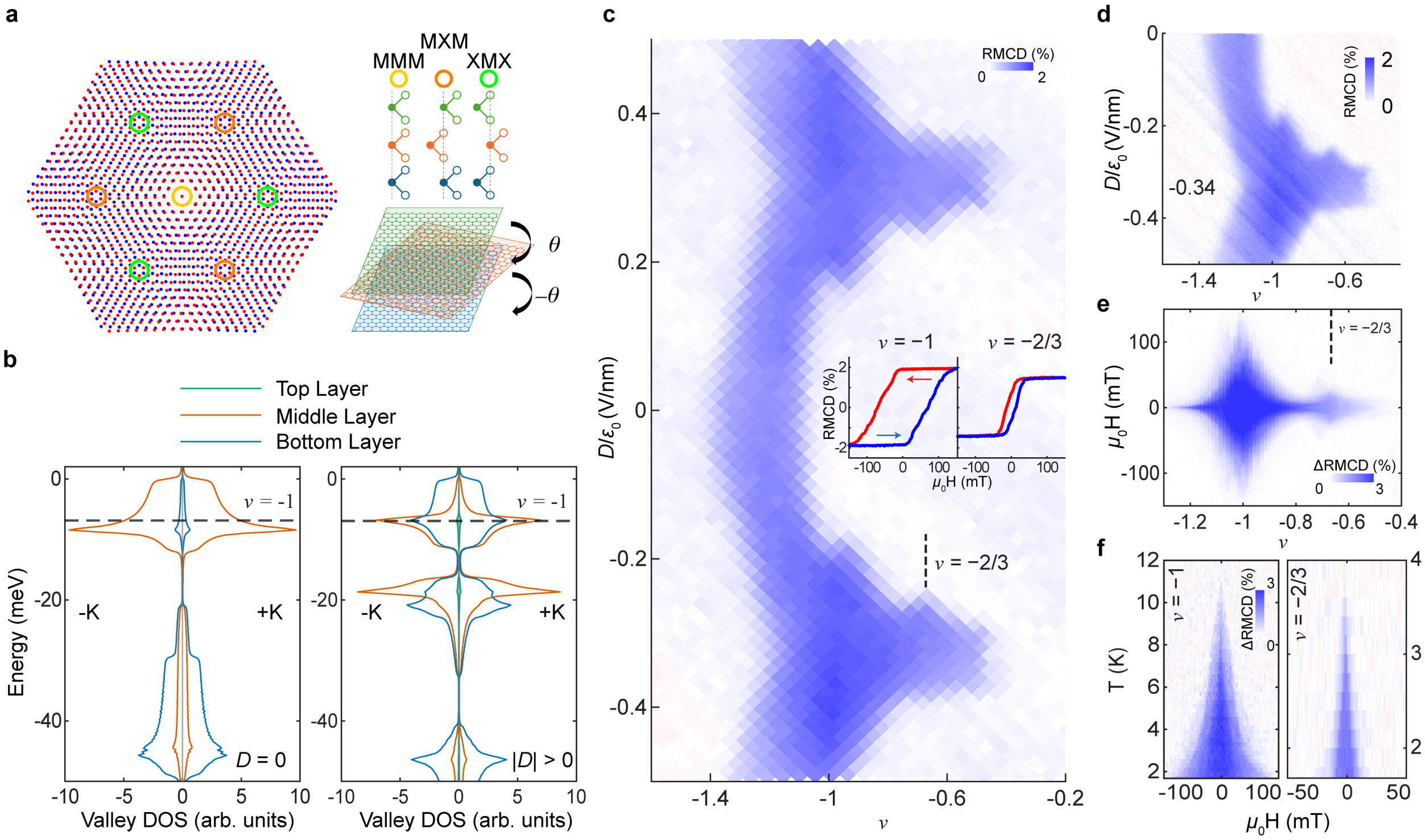

**Fig. 1 | Displacement field tunable moiré geometry and ferromagnetism. a,** Left: Moiré pattern in alternating-twisted trilayer $MoTe_2$ ($ATMoTe_2$). Blue and red dots represent the Mo atoms of the outer and inner layers, respectively. High symmetry points are marked. Top right: Side view of the atomic configurations at high symmetry points of the moiré lattice. M and X denote Mo and Te atoms, respectively. Bottom right: Stacking geometry of $ATMoTe_2$. **b,** Calculated layer-resolved K-valley density of states (DOS) as a function of energy for $|D| = 0$ (left panel) and $|D| > D_c$ (right panel). At high $D$, middle layer states hybridize with either the top or bottom layer, depending on the field direction. **c**, Reflective magnetic circular dichroism (RMCD) signal as a function of filling factor ($\nu$) and displacement field ($D$) for 3.9° twisted $ATMoTe_2$. Two regions of pronounced magnetism can be seen at $D/\epsilon_0 = \pm\, 0.34$ V/nm. Inset: RMCD signal versus $\mu_0 H$ swept back and forth at $\nu = -1$ and $-2/3$, respectively. **d,** Higher resolution plot of RMCD at negative $D$, showing an enhancement of critical displacement field $D$ at $\nu = -2/3$. **e,** Hysteretic component of RMCD (ΔRMCD) vs. $\nu$ and $\mu_0 H$ with fixed $D/\epsilon_0 = -0.34$ V/nm. Enhancement of coercive fields at $\nu = -1$ and $\nu = -2/3$ is observed. **f,** ΔRMCD versus temperature and $\mu_0 H$ at $\nu = -1$ (top) and $\nu = -2/3$ (bottom). The Curie temperatures are about 11 K and 3.5 K for $\nu = -1$ and $-2/3$, respectively. All data are taken at 1.6K except the temperature dependence in panel (f).

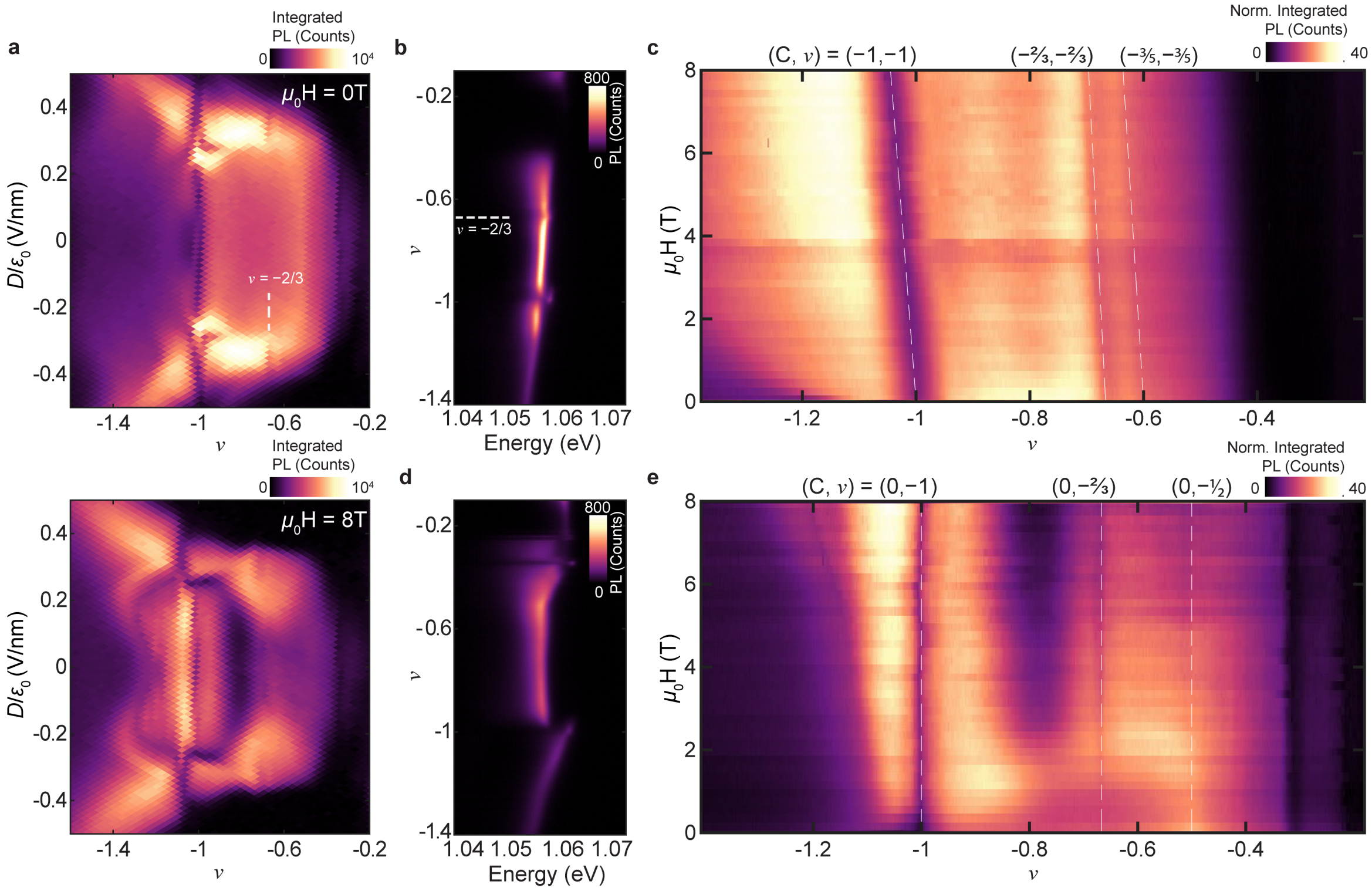

**Fig. 2 | Electrically tunable integer and fractional Chern insulators. a,** Spectrally integrated photoluminescence (PL) intensity as a function of $\nu$ and $D$. The top and bottom panels are taken at zero and 8T magnetic fields, respectively. **b,** PL intensity plots as a function of photon energy and filling factor ($\nu$) at $D/\epsilon_0 = -0.34$ V/nm. **c,** Spectrally integrated PL intensity versus $\mu_0 H$ and $\nu$ at $D/\epsilon_0 = -0.34$ V/nm. Dashed lines with Streda slope are indicated with corresponding filling and Chern numbers ($C$, $\nu$). **d,e,** Same as **b,c,** but for $D/\epsilon_0 = 0$ V/nm.

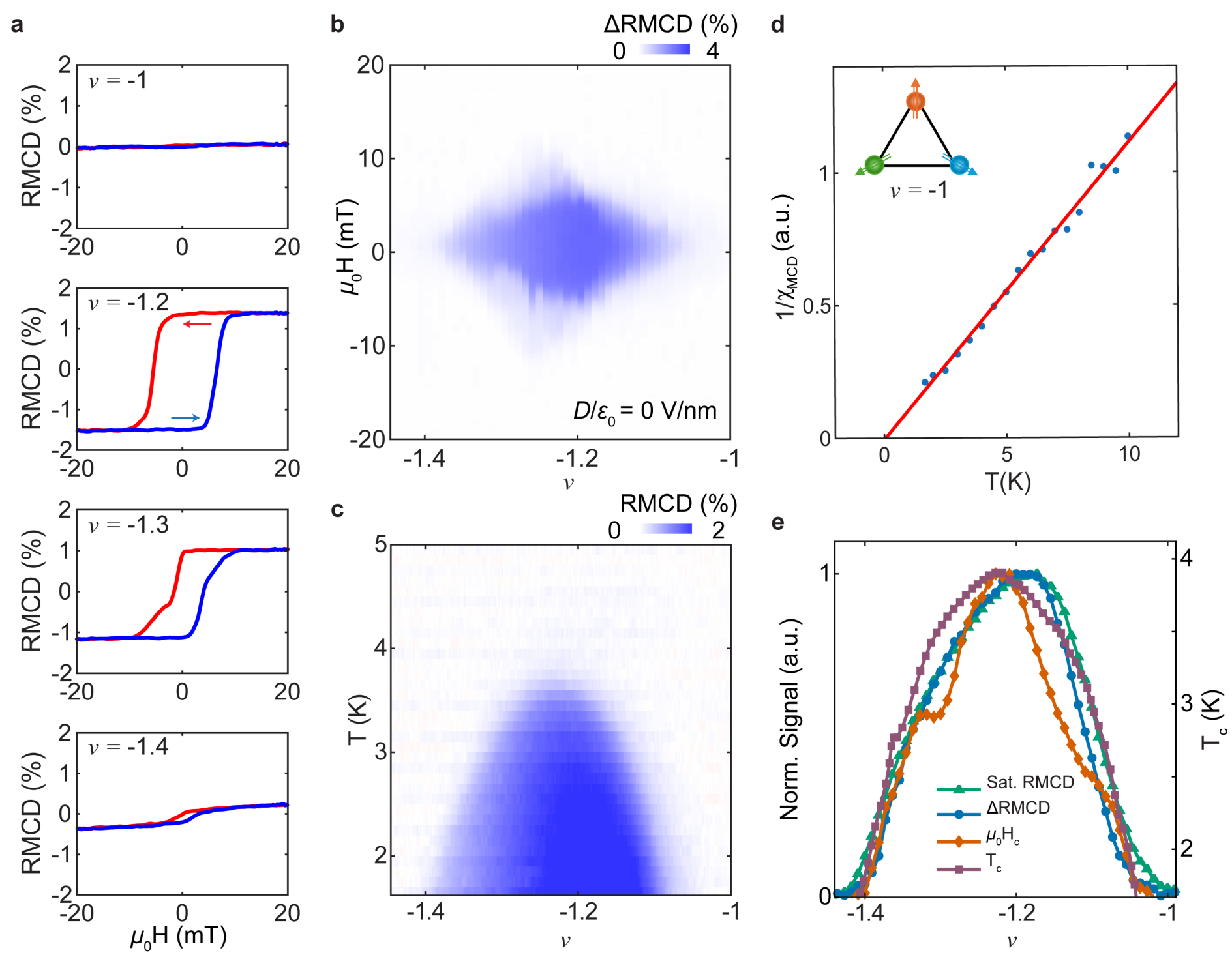

**Fig. 3 | Observation of double-exchange ferromagnetism in a doped triangular lattice.** All data are taken at $D/\epsilon_0 = 0$ V/nm. **a,** RCMD versus cycling magnetic fields at selected filling near $\nu = -1$. **b**, Hysteretic component of RMCD ($\Delta$RMCD) vs. $\nu$ and $\mu_0 H$, demonstrating ferromagnetism appearing for $\nu$ beyond $-1$, i.e. an over-doped correlated insulator. **c**, RMCD versus temperature as a function of $\nu$. The Curie temperature reaches maximum of around 3.8K near $\nu = -1.2$. Both (b) and (c) demonstrate that upon doping $\nu$ past $-1$, coercive fields and Curie temperature first increases, reaches a maximum, and then drops as filling increases. **d**, Curie-Weiss fitting of inverse RCMD slope near $\mu_0 H = 0$, an approximation of magnetic susceptibility $\chi$, at $\nu = -1$. The Curie-Weiss temperature is about zero, suggesting weak magnetic interaction. **e**, Curie temperature as a function of filling factor, overlaid with normalized coercive field, ΔRMCD at $\mu_0 H = 0$, and saturated RMCD. The collapsing of these four independent observables demonstrates an intrinsic single energy scale governing the ferromagnetism.

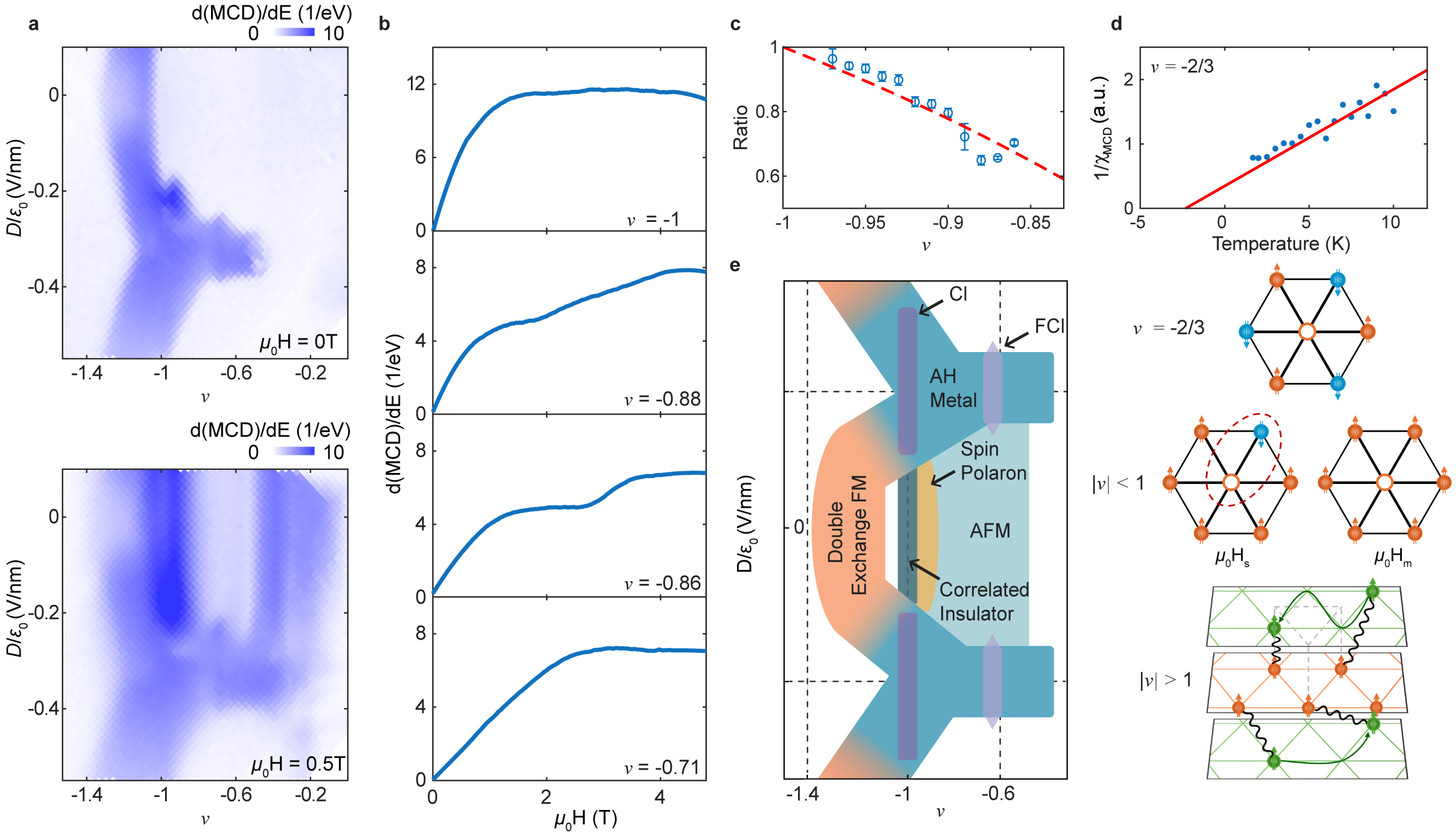

**Figure 4. Signatures of spin polaron and co-linear antiferromagnetism. a,** d(MCD)/dE as a function of $\nu$ and $D$ at $\mu_0 H = 0$ (top) and $\mu_0 H = 0.5$T (bottom). The comparison of the two reveals distinct magnetic interactions outside the ferromagnetic range, such as below $\nu = -1$ and $-1/2$. **b**, d(MCD)/dE vs. magnetic field at selected $\nu$ for $\nu$ below $-1$. The plateau behavior at intermediate magnetic fields suggests the formation of spin polarons. **c**, Extracted ratio of the d(MCD)/dE plateau to the fully saturated value, which follows $\frac{1-3|\delta|}{1-|\delta|}$, with $\delta = |\nu| - 1$. It is consistent with spin polaron formation. **d**, Inverse of the slope of RMCD near $\mu_0 H = 0$T, used as an approximation of magnetic susceptibility, as a function of temperature at $\nu = -2/3$. The Curie-Weiss fit yields a Curie-Weiss temperature near $-2$K, suggesting an antiferromagnetic state. **e**, Left: Schematic of the $\nu$ versus $D$ phase diagram for ATMoTe$_2$. Right: Schematic of the spin configurations at $D/\epsilon_0$ = 0 V/nm and various filling $\nu$. For $\nu = -2/3$ (top): schematic of a co-linear antiferromagnetic state. For $|\nu| < 1$ (middle): schematic of the spin polaron phase at saturation field ($\mu_0 H_s$) and magnetic transition field ($\mu_0 H_m$). For $|\nu| < 1$ (bottom): schematic of double-exchange ferromagnetism.

# Extended Data for

# Title: Electrically Programmable Correlated Topology and Magnetism in a Moiré Trilayer

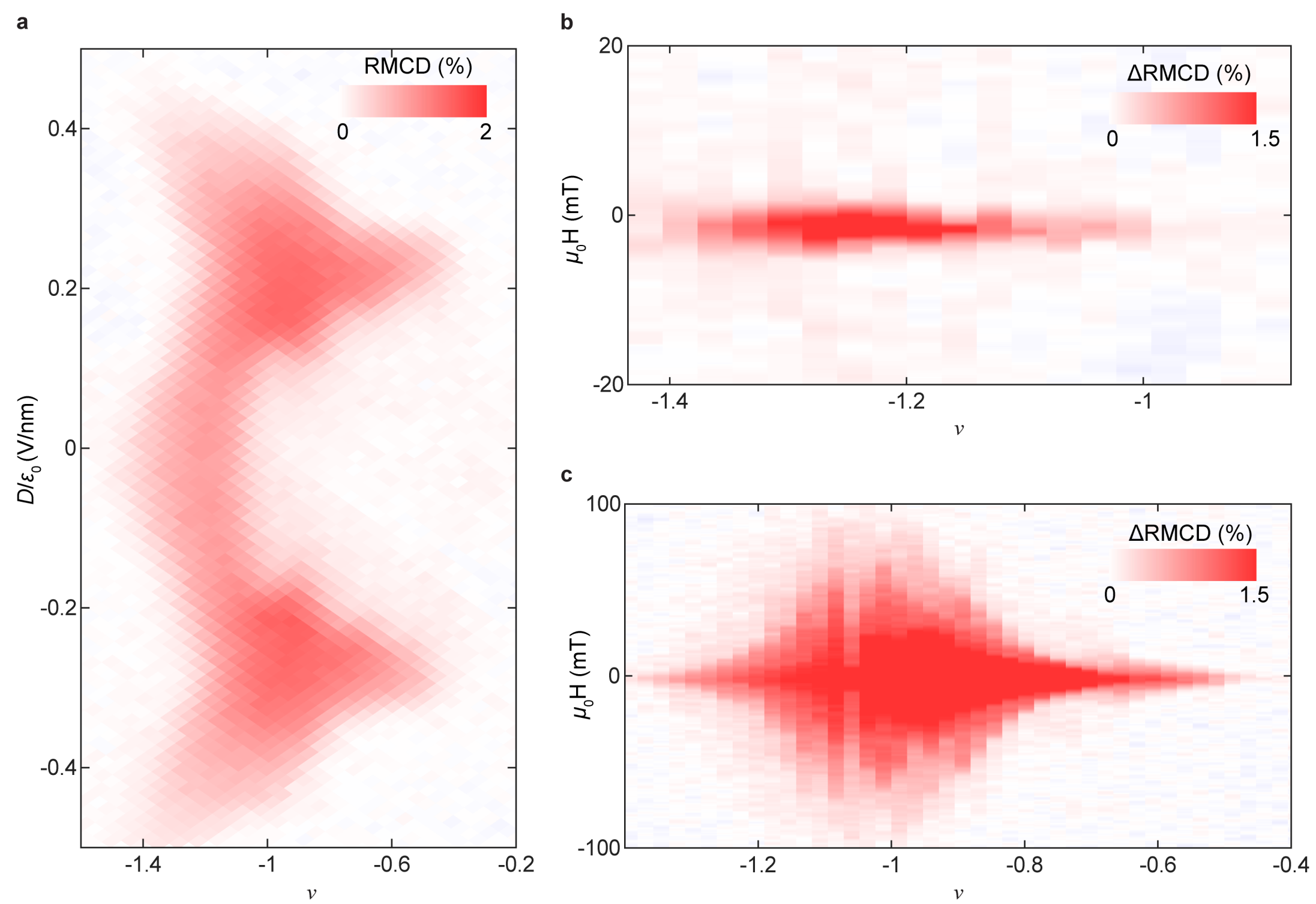

**Extended Fig. 1 | RMCD of another ATMoTe$_2$ device. a,** Reflective magnetic circular dichroism (RMCD) signal as a function of filling factor ($\nu$) and displacement field ($D$) for 3.6° alternating trilayer tMoTe$_2$. Two regions of enhanced magnetism can be seen at $D/\epsilon_0 = \pm\, 0.28$ V/nm, and finite magnetism can be seen at $D/\epsilon_0 = 0$ V/nm. **b,c,** Hysteretic component of RMCD (ΔRMCD) vs. $\nu$ and $\mu_0 H$ at a fixed $D/\epsilon_0 = 0$ V/nm (**b**) and $D/\epsilon_0 = -0.28$ V/nm (**c**).

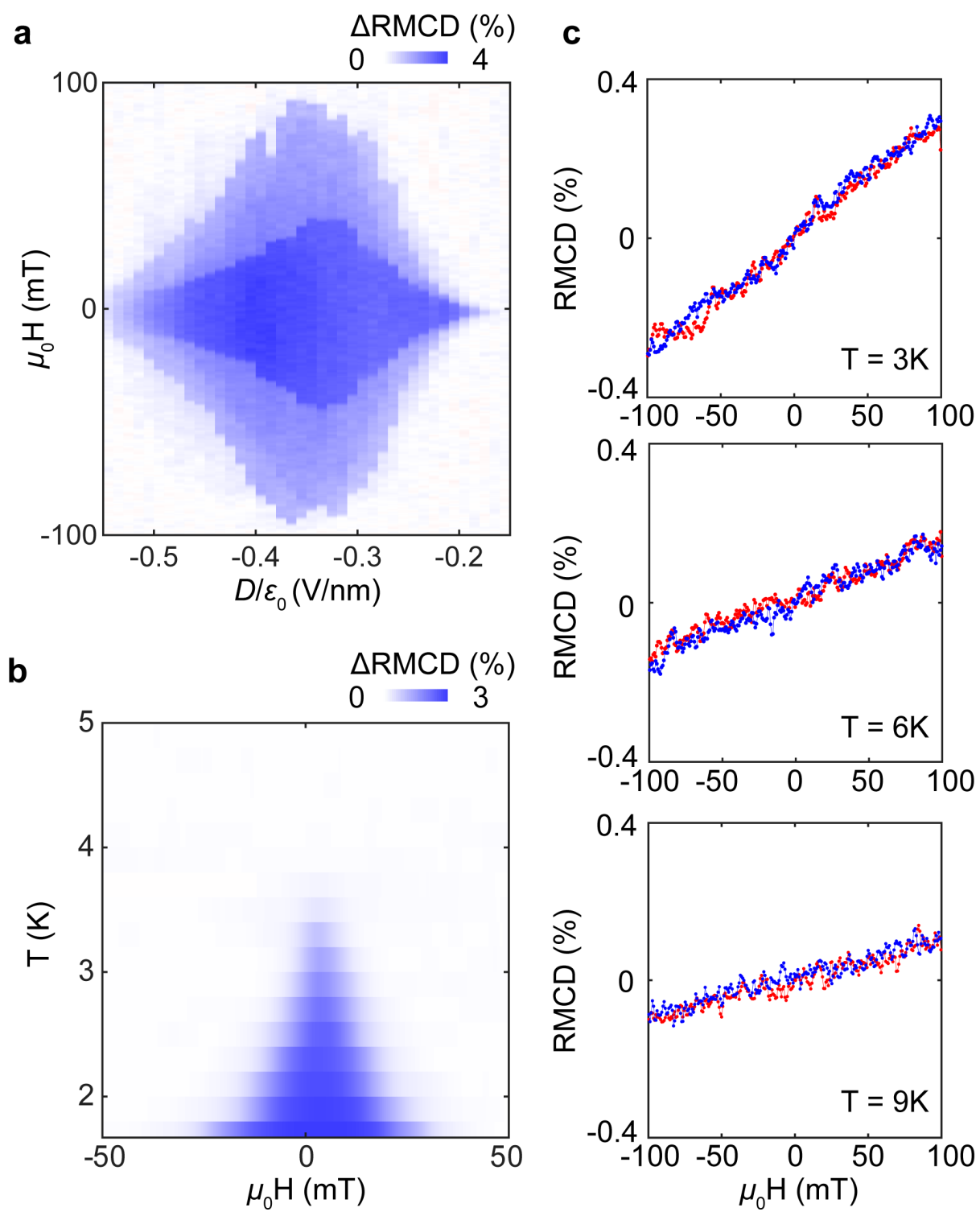

**Extended Fig. 2 | Displacement field and temperature dependence of magnetism around $\boldsymbol{\nu = -1}$.** **a,** ΔRMCD vs. $D$ and $\mu_0 H$ with fixed filling factor $\nu = -1$. ΔRMCD signal and $\mu_0 H_c$ decrease dramatically as applied $D$ field is decreased. **b,** ΔRMCD versus temperature and $\mu_0 H$ at $\nu = -1.2$ and $D/\epsilon_0 = 0$ V/nm. The Curie temperature is around 3.8K. **c,** Magnetic field dependence of RMCD around $\mu_0 H = 0$T for $\nu = -1$ and $D/\epsilon_0 = 0$ V/nm at T = 3K (top), T = 6K (middle) and T = 9K (bottom).

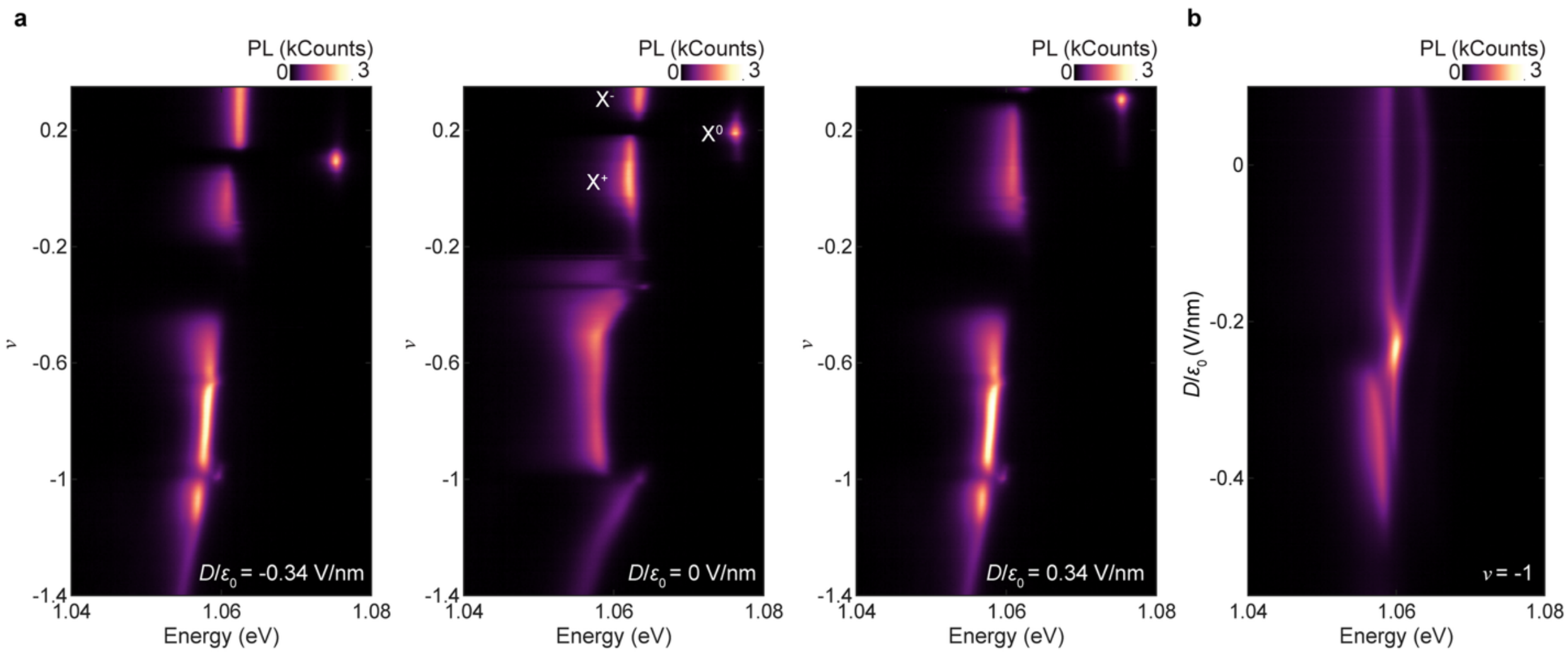

**Extended Fig. 3 | Doping and displacement field dependent photoluminescence (PL). a,** PL intensity plot as a function of photon energy and filling factor $\nu$ at $D/\epsilon_0$ = $\pm$0.34 V/nm (left, right) and $D/\epsilon_0$ = 0 V/nm (middle). $X_0$: neutral exciton; $X^+$: positively charged trion; $X^-$: negatively charged trion. **b,** PL intensity plot as a function of photon energy and displacement field $D$ at hole filling factor $\nu = -1$. The bright PL near $D/\epsilon_0 = -0.22$ V/nm indicates the gap closure associated with the topological phase transition.

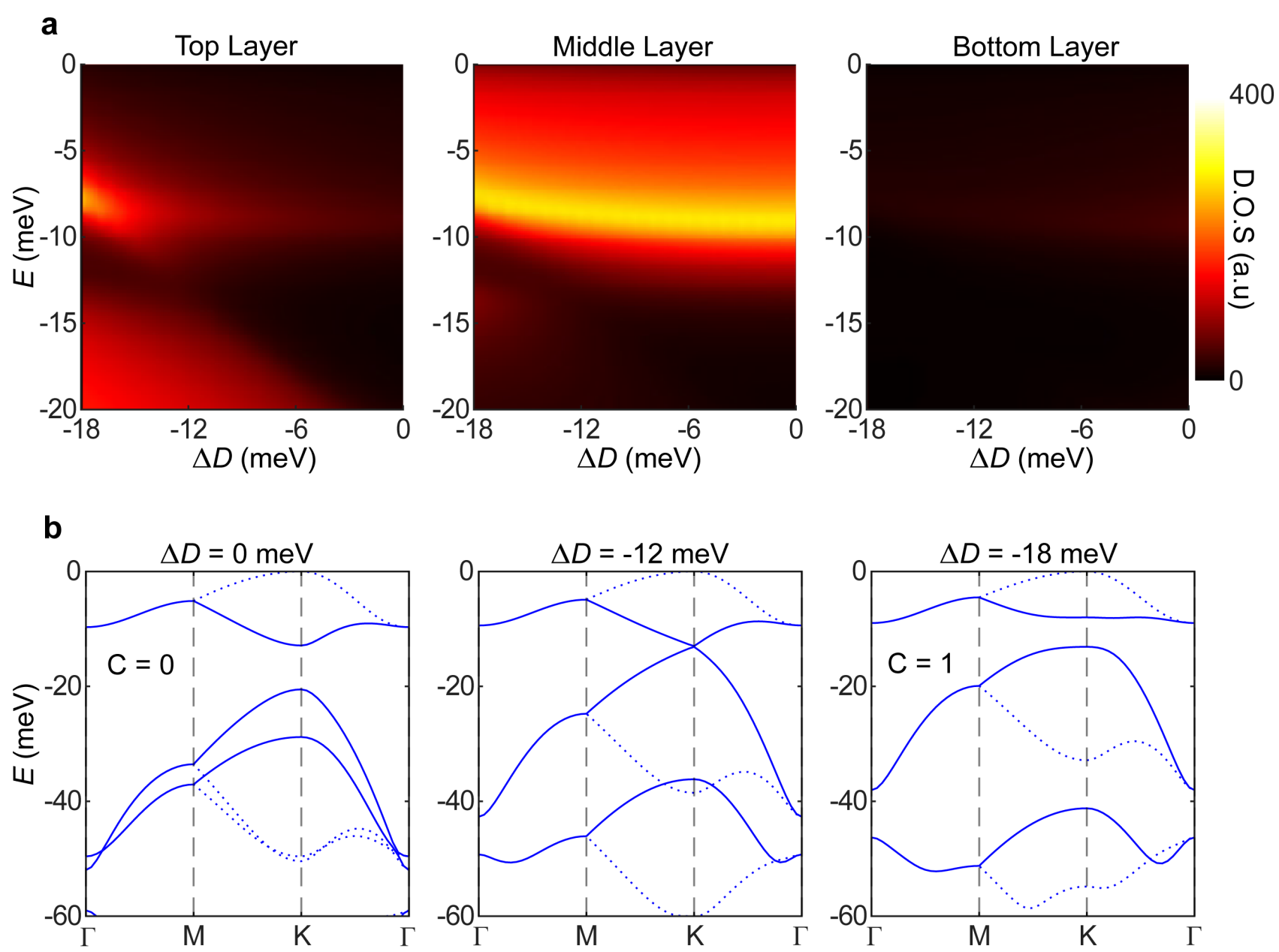

**Extended Fig. 4 | Theoretical calculations of ATMoTe₂. a,** Density of states calculations of the K valley of each $MoTe_2$ layer as a function of applied displacement field ($\Delta D$) and energy ($E$). Hybridization occurs between the top and middle layer with applied $\Delta D$. **b,** Band structure calculations of $ATMoTe_2$ at various $\Delta D$. A topological phase transition in the K valley can be observed at $\Delta D = -12$ meV.

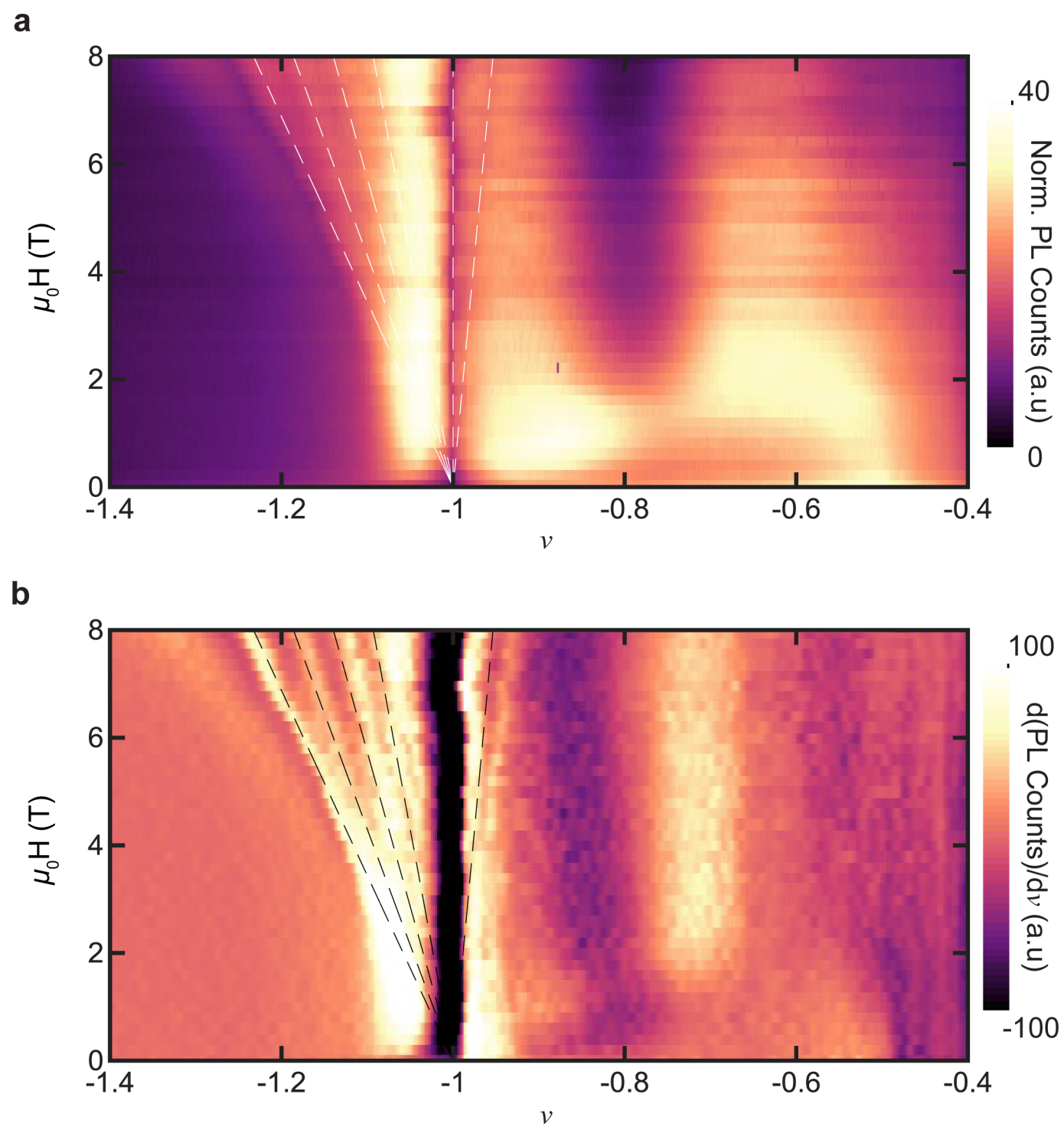

**Extended Fig. 5 | Landau level formation below critical D. a,** Spectrally integrated PL intensity versus $\mu_0 H$ and $\nu$ at $D/\epsilon_0 = -0.13$ V/nm. Dashed lines correspond to Landau level formation, with slopes of −5, −4, −3, −2 and +1 originating from $\nu = -1$. **b,** Derivative of PL counts with respect to filling factor $\nu$ of **a**.

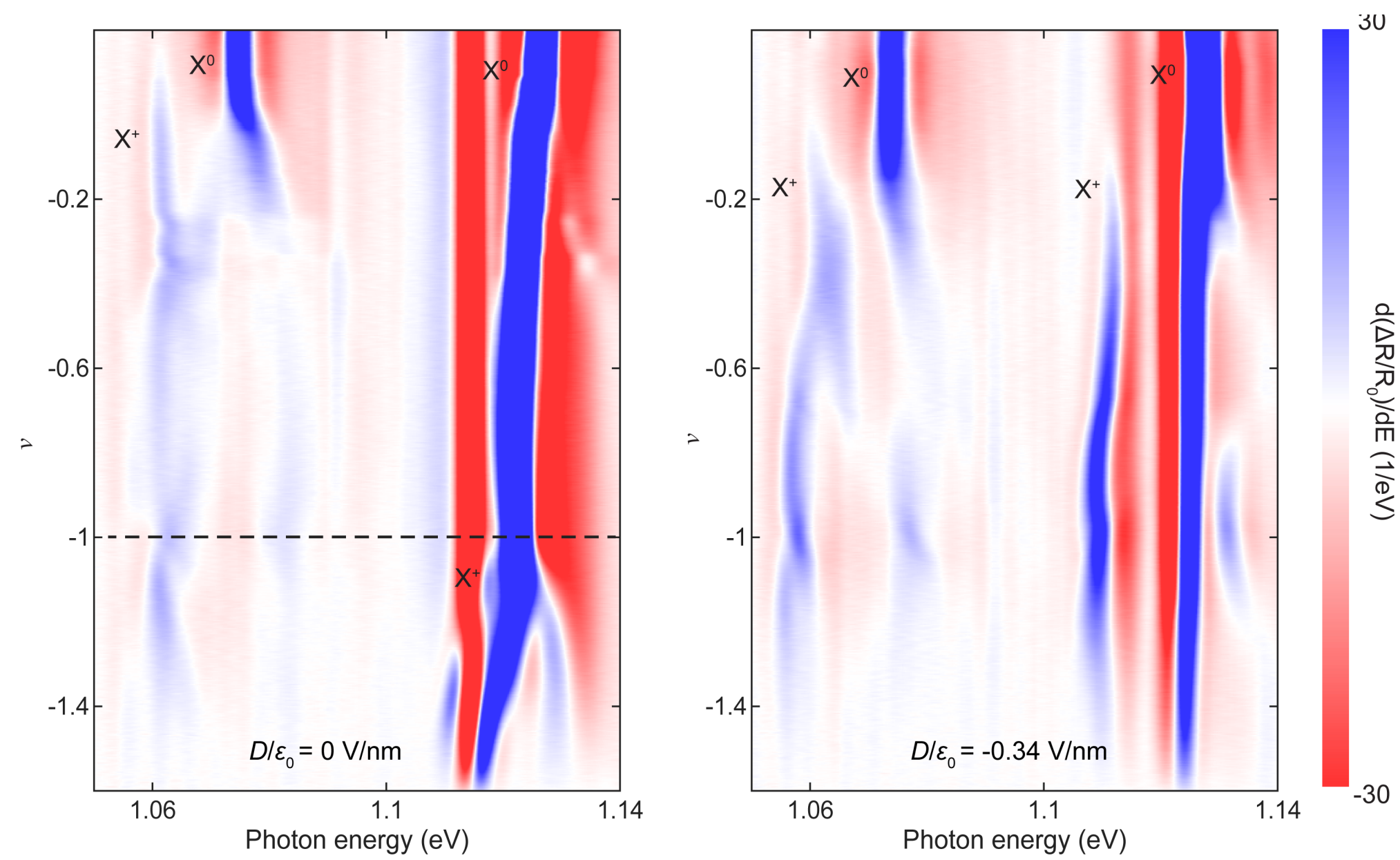

**Extended Fig. 6 | Reflectance spectrum of ATMoTe₂.** Derivative with respect to photon energy $E$ of white light reflectance spectrum as a function of filling factor $\nu$, at $D/\epsilon_0$ = 0 V/nm (left) and at $D/\epsilon_0 = -0.34$ V/nm (right). Compared with the PL data, we identify that the low energy resonances (~1.07 eV) are neutral excitons in the middle layer, while the high energy resonances (~1.13 eV) are neutral excitons in the top and bottom layer. At $D/\epsilon_0$ = 0 V/nm, doped charge carriers are initially introduced into the middle layer, forming redshifted trions. The charge neutral excitons in the top and bottom layer remain. At $\nu = -1$, charge carriers begin to dope the top and bottom layers, leading to a similar red shift in the high energy peak, indicating trion formation in the top and bottom layers. Conversely in the $D/\epsilon_0 = -0.34$ V/nm regime, charge carriers are simultaneously doped into both the middle and top layers, leading to simultaneous trion formation in both layers.

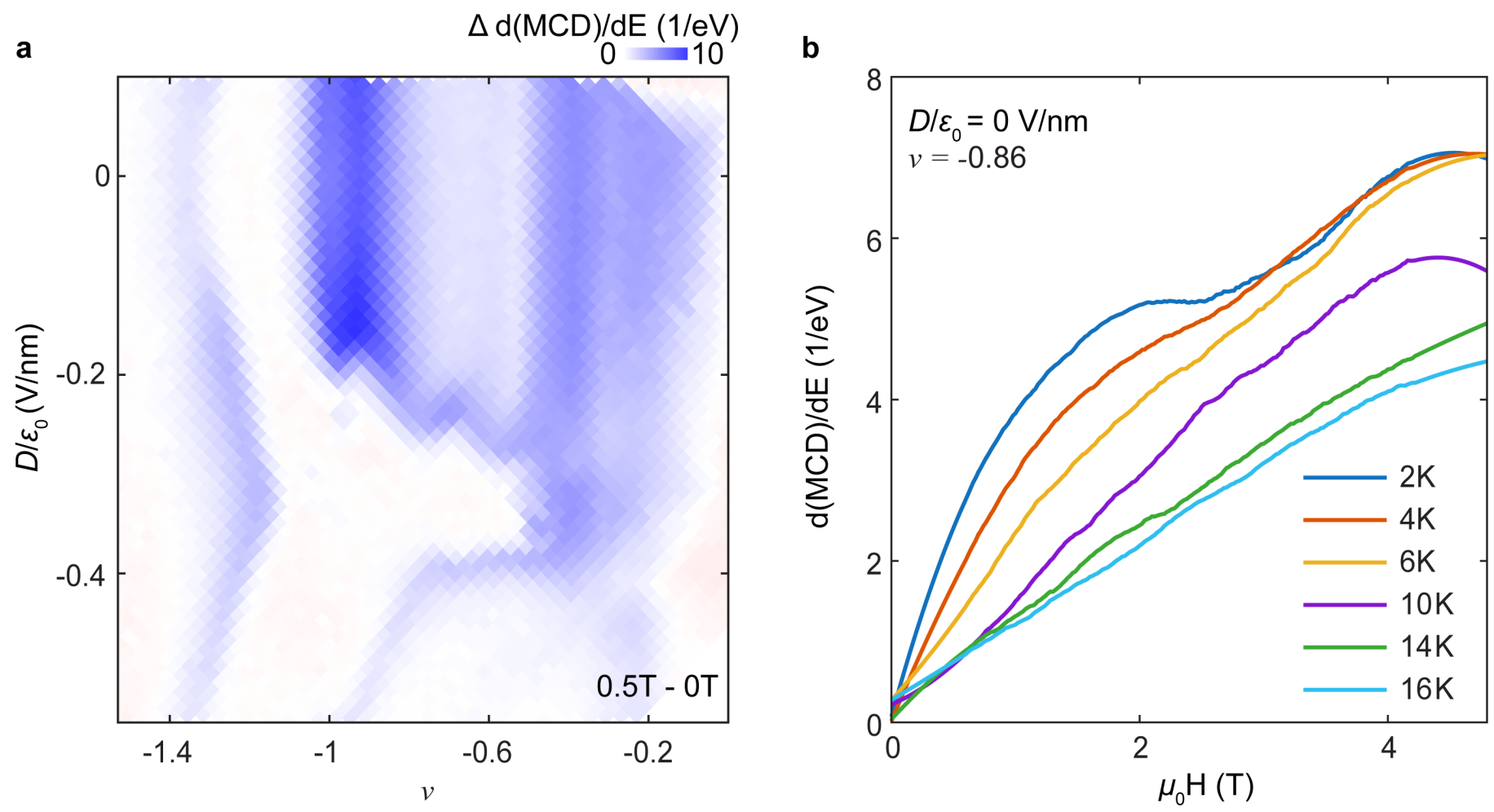

**Extended Fig. 7 | Magnetic evolution of non-ferromagnetic states. a,** Difference of d(MCD)/dE signals between the 0.5T and the 0T data in Main Figure 4a. The d(MCD)/dE signal in the ferromagnetic regime shows little change with applied magnetic field. **b**, Magnetic field dependence of d(MCD)/dE for $\nu = -0.86$ and $D/\epsilon_0$ = 0 V/nm at various temperatures.